\documentclass{emulateapj}

\usepackage[squaren]{SIunits}

\usepackage{ wasysym }
\usepackage{graphicx}
\usepackage{url}
\usepackage{epstopdf}
\newcommand {\HI}     {\ion{H}{1}}   %  HI
\newcommand {\HeI}    {\ion{He}{1}}   %  HeI
\newcommand {\HeII}  {\ion{He}{2}}    %  HeII
\shorttitle{AGN EUV Spectra}
\shortauthors{Tilton et al.}

\begin{document}
\defcitealias{nistasd}{NIST Atomic Spectra Database}

\renewcommand*{\thefootnote}{\fnsymbol{footnote}}
\title{\textit{HST}-COS Observations of AGNs. III.  Spectral Constraints in the Lyman Continuum from Composite COS/G140L Data\footnote{Based on observations made with the NASA/ESA \textit{Hubble Space Telescope}, obtained from the data archive at the Space Telescope Science Institute. STScI is operated by the Association of Universities for Research in Astronomy, Inc., under NASA contract NAS5-26555.}}

\author{Evan M. Tilton\altaffilmark{1}, Matthew L. Stevans\altaffilmark{2}, J. Michael Shull\altaffilmark{1}, and Charles W. Danforth\altaffilmark{1}}
%\affil{CASA, Department of Astrophysical and Planetary Sciences, University of Colorado, 389-UCB, Boulder, CO 80309, USA.}
\email{evan.tilton@colorado.edu, stevans@astro.as.utexas.edu,}
\email{michael.shull@colorado.edu, charles.danforth@colorado.edu}
\altaffiltext{1}{CASA, Department of Astrophysical and Planetary Sciences, University of Colorado, 389-UCB, Boulder, CO 80309, USA.}
\altaffiltext{2}{Department of Astronomy, The University of Texas at Austin, Austin, TX 78712, USA.}

\begin{abstract}
The rest-frame ultraviolet (UV) spectra of active galactic nuclei (AGNs) are important diagnostics of both  accretion disk physics  and  their contribution  to the metagalactic ionizing UV background. Though the mean AGN spectrum is well characterized with composite spectra at wavelengths greater than 912~\AA, the shorter-wavelength extreme-UV (EUV) remains poorly studied. In this third paper in a series on the spectra of AGNs, we combine 11 new spectra taken with the Cosmic Origins Spectrograph on the \textit{Hubble Space Telescope} with archival spectra to characterize the typical EUV spectral slope of AGNs from $\lambda_{\rm rest}\sim \unit{850}{\angstrom}$ down to $\lambda_{\rm rest}\sim\unit{425}{\angstrom}$. Parameterizing this slope as a power law, we obtain $F_\nu\propto \nu^{ -0.72\pm 0.26}$, but we also discuss the limitations and systematic uncertainties of this model. We  identify broad  emission features in this spectral region, including emission due to ions of O, Ne, Mg, and other species, and we limit the intrinsic \HeI\ 504~\AA\ photoelectric absorption edge opacity to $\tau_{\rm HeI}<0.047$.
\end{abstract}


\defcitealias{shu12a}{Paper I}
\defcitealias{ste14}{Paper II}

%Hello @OverheardOnAph - I hope you are doing well! You haven't posted much lately!

\section{INTRODUCTION}
\renewcommand*{\thefootnote}{\arabic{footnote}}
\setcounter{footnote}{0}
Active galactic nuclei (AGNs) represent the phase of rapid gas accretion and the associated high-energy emission around the supermassive black holes at the centers of their host galaxies. Their spectral properties in the ultraviolet (UV) and optical wavelength regimes are complex and diverse. Most commonly, their extreme-UV (EUV;  $\lambda<912$~\AA) and far-UV (FUV; $\unit{1000}{\angstrom}<\lambda<\unit{2000}{\angstrom}$) spectra consist of broad ( $>$\unit{1000}{\kilo\meter~\reciprocal\second}) and narrow emission lines superimposed on top of a roughly power-law continuum spectrum thought to arise in the AGN's accretion disk \citep[e.g.,][]{kro99}. As some of the most luminous objects in the universe, AGNs are important to a wide array of astrophysical phenomena at all scales. Understanding their spectra is therefore crucial to understanding a variety of processes.

At the smallest scales, AGN spectra probe the structure of the accretion processes that produce the continuum flux and ultimately control the flow of gas onto the black hole. This flux in turn regulates the ionization conditions of the gas that forms the broad emission lines. The kinematics and geometry of the broad emission line region itself remain poorly understood \citep[e.g.,][for a review]{gas09},  but the spectral properties of these lines serve as the primary diagnostics of the masses of the AGN black holes \citep[e.g.,][]{pet93,ves06,til13}. At galactic scales, we are just beginning to understand the broader role of AGN mechanical and radiative luminosity in regulating processes such as star formation in their host galaxies \citep[e.g.,][and references therein]{hop09,sil09,ara13}.

On the largest cosmological scales, radiation from AGNs governs much of the evolution of the intergalactic medium (IGM), the primary reservoir of baryons at all redshifts. Because galactic stellar populations cannot produce sufficient radiation to reionize \HeII, AGNs are thought to be the primary cause of \HeII\ reionization at $z\approx 3$ \citep[e.g.,][]{shu04}. At lower redshifts ($z<3$),  AGNs dominate the UV metagalactic ionization background radiation \citep[UVB;][and references therein]{haa12}, which largely controls the ionization and thermal state of the IGM. Low-redshift UV surveys of IGM absorption lines \citep[e.g.,][]{dan08,tri08,tho08,til12,dan14} have now characterized the distribution of \HI\ and metal absorbers. However, the IGM baryon census remains incomplete \citep{shu12b} and sensitive to assumptions about the metagalactic ionizing background, which in turn depends on an understanding of the mean properties of the UV spectra of AGNs. Recently, there have been doubts raised in the literature \citep{kol14,shu15,kha15} about the ability of current estimates of the UVB to match observational results in the low-redshift Ly$\alpha$ forest of \HI\ absorbers.

A common technique for constraining the average behavior of AGN spectra is the construction of composite spectra from  individual spectra of different AGNs. Though such a technique cannot address the dramatic variations among AGN spectra, it can constrain the consistent mean behavior of their spectral energy distributions while also allowing the detection of spectral features that may be too weak to detect in any one observation, such as faint emission lines or absorption edges. The rest-frame FUV continuum has been well-studied in composites that draw from samples of moderate-to-high redshift AGNs, where their rest-frame UV is accessible to ground-based telescopes \citep{fra91,car99,bro01,van01,xie15,sel15}. Probing deeper into the rest-frame EUV from the ground, however, is complicated by the increased density of IGM absorbers at higher redshifts. Further, the EUV and FUV spectra of low-redshift AGNs are entirely inaccessible from ground-based facilities.

For these reasons, the best probes of the rest-frame EUV and FUV continua come from space-based observations in the UV. Various authors have constructed EUV and FUV composites  using the Faint Object Spectrograph (FOS), the Goddard High-Resolution Spectrograph (GHRS), the Space Telescope Imaging Spectrograph (STIS), and the Wide Field Camera~3 (WFC3) UVIS grism onboard the \textit{Hubble Space Telescope} \citep[\textit{HST};][]{zhe97,tel02,lus15}, and others have used the \textit{Far-Ultraviolet Spectroscopic Explorer} \citep[\textit{FUSE};][]{sco04}. Parameterizing the EUV continua as power laws, $F_\lambda\propto\lambda^{\alpha_\lambda}$ and $F_\nu\propto\nu^{\alpha_\nu}$ with $\alpha_\nu=-(2+\alpha_\lambda)$, these authors arrived at  spectral indices that differed owing to systematic differences in target selection, composite construction, and the separation of continuum and emission-line flux, which can be complicated by broad emission lines such as \ion{Ne}{8} $\lambda\lambda 770, 780$. In particular, \citet{sco04} obtained a substantially harder EUV spectrum, $\alpha_\nu=-0.56^{+0.38}_{-0.28}$, than \citet[][$\alpha_\nu=-1.57\pm 0.17$ in a radio-quiet subsample]{tel02} or \citet[][$\alpha_\nu=-1.70 \pm 0.61$]{lus15}.

This paper is the third in a series that use the Cosmic Origins Spectrograph (COS) on \textit{HST} to constrain the EUV  and FUV spectra of AGN through composite spectra. The improved sensitivity of COS compared to past UV spectrographs makes it an ideal instrument for studying a large sample of AGN at high signal-to-noise ratios (S/N). In \citet[][hereafter \citetalias{shu12a}]{shu12a} we presented a pilot study of composite construction using 22 AGNs observed with the medium-resolution grating modes of COS (G130M and G160M). \citet[][hereafter \citetalias{ste14}]{ste14} extended this work, constructing a medium-resolution COS composite of the rest-frame FUV and EUV from 159 AGNs at redshifts $0.001 < z_{\rm AGN} < 1.476$. This study obtained an EUV slope $\alpha_\nu=-1.41 \pm 0.15$, consistent with values in \citetalias{shu12a}, \citet{tel02}, and \citet{lus15}. Unfortunately, the COS archive of spectra from which \citetalias{ste14} drew had fewer than 10 AGNs sampling $\lambda<\unit{600}{\angstrom}$ and none at $\lambda<\unit{480}{\angstrom}$.

In this paper, we present 11 new, low-resolution COS/G140L spectra of AGNs at $1.45\leq z_{\rm AGN}\leq 2.14$ from \textit{HST} Cycle 21 observations (PID 13302, Shull P.I.). We combine them with existing medium-resolution data to construct composite spectra that extend to shorter rest-frame wavelengths, $\unit{400}{\angstrom}<\lambda<\unit{850}{\angstrom}$, more robustly than our previous COS composites. This paper updates the work presented in \citetalias{shu12a} and \citetalias{ste14} and improves the constraints and statistical bounds on the spectral behavior of the shortest wavelengths probed by \citetalias{ste14}. In Section~\ref{sec:meth} we describe the sample, data processing, and composite contruction techniques. In Section~\ref{sec:disc}, we discuss our results for the typical EUV slope, limits on the \HeI\ \unit{504}{\angstrom} photoelectric edge, applications to accretion disk models, and the broad emission lines observed in the composite. Finally, in Section~\ref{sec:conc} we summarize our results.

\section{Methodology}
\label{sec:meth}
At a general level, our analysis is similar to the methods described in \citetalias{ste14}. However, a number of small differences arise, owing to our use of lower resolution COS/G140L data and our focus on a smaller, EUV wavelength range. We therefore describe our methodology  in this section. Throughout this paper we adopt atomic transition wavelengths and other atomic properties from \citet{mor03}. When that catalog does not contain the line of interest, we use the NIST Atomic Spectra Database.\footnote{\url{http://www.nist.gov/pml/data/asd.cfm}} Where necessary, we adopt a flat $\Lambda$CDM cosmology with $H_0=71~{\rm km~s^{-1}~Mpc^{-1}}$, $\Omega_m=0.27$, and $\Omega_\Lambda=0.73$. All neutral hydrogen column densities $N_{\rm HI}$ are expressed in units of ${\rm cm^{-2}}$.

\subsection{Sample, Data Acquisition, and Processing}
\label{sec:targmeth}
%these two tabs finalized%
\begin{deluxetable*}{lcccrcc}
\tabletypesize{\tiny}
\tablecaption{\textit{HST}/COS Observation Details\label{tab:obstab}}
\tablecolumns{7}
\tablewidth{0pt}
\tablehead{
    \colhead{Target} & \colhead{Right Ascension} & \colhead{Declination} & \colhead{Grating} & \colhead{Exp. Time} & \colhead{Obs. Date} & \colhead{Program ID}\\
    \colhead{} & \colhead{(J2000)} & \colhead{(J2000)} & & \colhead{(s)} & \colhead{(GMT)} & \
}
\startdata
\cutinhead{COS/G140L}
              HS1803+5425 & 18 04 37.478 & +54 25 40.65 & G140L & 5460.736 & 2014-02-26  & 13302\\
              HE1120+0154 & 11 23 20.720 & +01 37 47.56 & G140L & 4992.768 & 2014-06-19  & 13302 \\
              SBS1010+535 & 10 13 30.165 & +53 15 59.71 & G140L & 10921.440 & 2014-01-30, 2013-10-01 & 13302 \\
              HE0248-3628 & 02 50 55.326 & -36 16 35.76 & G140L & 9597.568 & 2014-06-28, 2014-12-06  & 13302 \\
                   US2504 & 11 29 50.178 & +26 52 53.97 & G140L & 5040.704 & 2015-02-15  & 13302 \\
  SDSSJ125140.83+080718.4 & 12 51 40.810 & +08 07 18.19 & G140L & 5000.800 & 2014-06-19 & 13302\\
  SDSSJ083850.15+261105.4 &08 38 50.153 &+26 11 05.26 & G140L & 5040.672 &2013-12-01 & 13302 \\
  SDSSJ094209.14+520714.5 & 09 42 09.163 & +52 07 14.54  & G140L& 5460.704 & 2013-10-06 & 13302\\
             PG1115+080A1 & 11 18 16.930 & +07 45 58.52 & G140L & 5000.704 &2014-02-12 & 13302\\
            HB89-1621+392 & 16 23 07.665 & +39 09 32.18 & G140L & 5108.800 &2013-10-11 & 13302 \\
              SBS1307+462 & 13 10 11.620 & +46 01 24.50 & G140L & 5270.720 & 2013-11-17  & 13302\\
\cutinhead{COS/G130M and COS/G160M}
  SDSSJ084349.49+411741.6 & 08 43 49.476 & +41 17 41.64 & G130M &4361.408 & 2011-05-11  & 12248\\
               &  &  & G160M& 7010.368 & 2011-05-11   & 12248\\
              HE0439-5254  &04 40 11.900  & -52 48 18.00 & G130M &8402.944  & 2010-06-10 & 11520\\
               &  &  & G160M& 8936.064&  2010-06-10    & 11520\\
  SDSSJ100535.24+013445.7 & 10 05 35.257 & +01 34 45.60 & G130M  & 11251.740 & 2011-03-13 & 12264\\
               &  &  & G160M& 22413.152 & 2011-03-13   & 12264\\
               PG1206+459 & 12 08 58.011 & +45 40 35.48 & G130M &17361.120  &   	2010-01-04 & 11741\\
               &  &  & G160M&36093.824 & 	2010-01-04    & 11741\\
               PG1338+416  &13 41 00.780  & +41 23 14.10 & G130M & 22743.488 & 2010-05-30, 2010-05-24 & 11741\\
               &  &  & G160M& 35013.312 & 2010-05-24, 2010-05-25, 2010-05-27   & 11741\\
            LBQS1435-0134 & 14 37 48.284 & -01 47 10.78 & G130M & 22347.52 &2010-08-08, 2010-08-18   & 11741\\
               &  &  & G160M& 34165.984& 2010-08-18, 2010-08-17,  2010-08-22   & 11741\\
               PG1522+101  & 15 24 24.580 & +09 58 29.70 &G130M  &16401.184  & 2010-09-12, 2010-09-14 & 11741 	\\
               &  &  & G160M& 23026.304& 2010-09-14,  	2010-09-07   & 11741 	\\
                Q0232-042  & 02 35 07.385 & -04 02 05.67 &G130M  & 16019.104 &  	2010-02-16, 2010-02-20 & 11741\\
               &  &  & G160M&22840.192 & 2010-02-20, 2010-01-09 & 11741\\
               PG1630+377  & 16 32 01.120 &+37 37 50.00  &G130M  & 22988.992 & 2009-12-13, 2010-11-25 & 11741\\
               &  &  & G160M&14304.032  &2010-08-01, 2010-11-26   & 11741\\
\enddata
\end{deluxetable*}

\begin{deluxetable*}{lllrrr}
\tabletypesize{\tiny}
\tablecaption{AGN Properties\tablenotemark{a}\label{tab:targs}}
\tablecolumns{6}
\tablewidth{0pt}
\tablehead{
    \colhead{Target} & \colhead{Redshift ($z_{\rm AGN}$)} & \colhead{$z_{\rm AGN}$ Reference} & \colhead{$\alpha_\lambda$} & \colhead{$F_{\rm 0}$} & \colhead{$\log\left[\frac{\lambda L_\lambda ({\rm 573~\AA})}{\rm erg~s^{-1}}\right]$} \
}
\startdata
\cutinhead{COS/G140L}
              HS1803+5425 & $1.448$ & \citet{eng98} & $-0.12\pm0.03$ & $1.9642\pm0.0555$ & $46.329\pm0.019$ \\
              HE1120+0154 & $1.471700\pm0.000438$ & \citet{hew10} & $-1.24\pm0.05$ & $1.1386\pm0.0241$ & $46.499\pm0.021$ \\
              SBS1010+535 & $1.515848\pm0.000444$ & \citet{hew10} & $-4.17\pm0.01$ & $0.0911\pm0.0005$ & $46.280\pm0.004$ \\
              HE0248-3628 & $1.536$ & \citet{wis00} & $-1.54\pm0.07$ & $2.2393\pm0.0602$ & $46.773\pm0.010$ \\
                   US2504 & $1.542621\pm0.000499$ & \citet{hew10} & $-1.88\pm0.02$ & $0.6622\pm0.0104$ & $46.421\pm0.011$ \\
  SDSSJ125140.83+080718.4 & $1.596100\pm0.005000$ & \citet{sch10} & $-1.40\pm0.03$ & $0.7683\pm0.0139$ & $46.397\pm0.016$ \\
  SDSSJ083850.15+261105.4 & $1.618279\pm0.000438$ & \citet{hew10} & $-1.80\pm0.07$ & $0.6582\pm0.0185$ & $46.458\pm0.026$ \\
  SDSSJ094209.14+520714.5 & $1.652971\pm0.000493$ & \citet{hew10} & $-2.53\pm0.03$ & $0.2592\pm0.0033$ & $46.298\pm0.024$ \\
             PG1115+080A1 & $1.735512\pm0.000439$ & \citet{hew10} & $-1.11\pm0.07$ & $0.6563\pm0.0201$ & $46.442\pm0.021$ \\
            HB89-1621+392 & $1.981361\pm0.000450$ & \citet{hew10} & $-1.52\pm0.01$ & $0.3133\pm0.0022$ & $46.274\pm0.010$ \\
              SBS1307+462 & $2.142306\pm0.000247$ & \citet{hew10} & $-2.25\pm0.07$ & $0.3204\pm0.0139$ & $46.565\pm0.008$ \\
\cutinhead{COS/G130M and COS/G160M}
  SDSSJ084349.49+411741.6 & $0.990788\pm0.000450$ & \citet{hew10} & $-1.54\pm0.07$ & $0.8770\pm0.0245$ & $45.875\pm0.033$ \\
              HE0439-5254 & $1.053$ & \citet{wis00} & $-1.34\pm0.02$ & $2.0865\pm0.0205$ & $46.239\pm0.008$ \\
  SDSSJ100535.24+013445.7 & $1.080900\pm0.000379$ & \citet{hew10} & $-1.61\pm0.20$ & $1.5900\pm0.0757$ & $46.228\pm0.060$ \\
               PG1206+459 & $1.164941\pm0.000436$ & \citet{hew10} & $-0.84\pm0.28$ & $3.3995\pm0.7101$ & $46.518\pm0.057$ \\
               PG1338+416 & $1.217076\pm0.000445$ & \citet{hew10} & $-2.07\pm0.03$ & $0.5149\pm0.0059$ & $46.095\pm0.007$ \\
            LBQS1435-0134 & $1.310790\pm0.000437$ & \citet{hew10} & $-0.67\pm0.06$ & $4.2471\pm0.0975$ & $46.678\pm0.028$ \\
               PG1522+101 & $1.328005\pm0.000445$ & \citet{hew10} & $-0.66\pm0.07$ & $4.1178\pm0.1478$ & $46.673\pm0.029$ \\
                Q0232-042 & $1.437368\pm0.000183$ & \citet{hew10} & $-0.95\pm0.09$ & $1.5086\pm0.0715$ & $46.423\pm0.084$ \\
               PG1630+377 & $1.478949\pm0.000439$ & \citet{hew10} & $-1.69\pm0.09$ & $2.4274\pm0.1901$ & $46.836\pm0.045$ \\
\enddata
\tablenotetext{a}{Power-law parameters are fit to the form $F_\lambda=F_{\rm 0}(\lambda /\unit{1100}{\angstrom})^{\alpha_\lambda}$ in the rest frame. Fluxes are in units of $10^{-15}{\rm~erg~cm^{-2}~s^{-1}~\AA^{-1}}$. Monochromatic luminosities, $\lambda L_\lambda$, are measured from luminosity distances and the spline fit at rest-frame 573~\AA\ (the normalization window used for the composite). Power-law parameters and luminosities include corrections for identified HI absorption, Galactic extinction, and resolution effects, as described in Section~2.1.}
\end{deluxetable*}

We observed 11 AGNs at $1.45\leq z_{\rm AGN}\leq 2.14$ with COS/G140L.  The COS/G140L grating provides a resolving power of 1500 to 4000 over a waveband of approximately 1100~\AA\ to 2150~\AA\ with the 1105~\AA\ central wavelength setting  used for all  exposures. These targets were selected on the basis of their high \textit{GALEX} fluxes \citep{bia14}, so the sample likely probes the high end of the AGN UV luminosity function. It is not a complete or unbiased sample, except in the sense that these targets represent the best COS observations of known bright AGNs in this redshift regime and were chosen without prior knowledge of their EUV spectral shapes. One other target in the redshift range of interest, Ton 34, has existing high-S/N observations with COS/G140L and other UV instruments available in the HST archives. We exclude this target from our analysis because it was observed owing to its position as an extreme outlier in the distribution of EUV slopes as well as its prominent broad absorption lines \citep{bin08a,bin08b,kro10}. It would therefore strongly bias the slope of our composite spectrum. In addition to the COS/G140L data, we also use almost all (9 of 10) medium-resolution targets from \citetalias{ste14} with $z_{\rm AGN}>0.98$, the redshift threshold above which all of our targets can be normalized to the same continuum window. The one exception is FIRST~J020930.7-043826 ($z_{\rm AGN}=1.131$), which we exclude because \HI\ continuum absorption at $z_{\rm LLS}=0.39035$ makes most of the short-wavelength end of the spectrum unusable. All observations are summarized in Table~\ref{tab:obstab}. Table~\ref{tab:targs} lists basic physical parameters of the targets, including redshifts retrieved from the literature and measured monochromatic luminosities at rest-frame 573~\AA\ (a wavelength region for which all targets have useable coverage). For redshifts without a reported formal uncertainty, we conservatively assume $\sigma_{z_{\rm AGN}}=0.01$. All luminosity measurements are taken from spline fits and include corrections for Galactic extinction, identified \HI\ absorption, and resolution effects (see below). They do not account for errors due to unidentified \HI\ absorption from intervening absorbers that are redshifted out of the observational band or the absolute flux calibration of the instrument.

All calibrated exposures for each target were coadded using \textsc{Coadd{\_}X1D} version 3.3 from the COS Tools website,\footnote{\url{http://casa.colorado.edu/~danforth/science/cos/costools.html}} described in detail by \citet{dan10} and \citet{kee12}. This version of the software implements several minor changes, such as improved exposure cross-correlation and scaling between gratings, compared to the version used in \citetalias{ste14}. Because the COS detector oversamples the line-spread function by $\sim 6$ pixels, the spectra were binned by three pixels during the coaddition process to improve the S/N per bin without degrading the resolution of the spectra. 

%alternate form of resfix: $C=\left( 0.961\pm 0.001\right) z_{\rm Ly\alpha}^{\left( -0.0045\pm 0.0005\right)}$ over roughly $z_{\rm Ly\alpha}<0.47$

We fit each spectrum with a combination of splines and polynomials using the same procedure as in \citetalias{shu12a} and \citetalias{ste14}. This process allows us to smooth and reconstruct the shape of the intrinsic AGN spectrum, removing intervening absorption lines from the IGM and interstellar medium (ISM) by interpolating over them, in contrast to other composites \citep[e.g.,][]{tel02,lus15} that apply statistical corrections for such absorption. However, the reduced resolving power of the G140L grating compared to the medium-resolution COS gratings leads to many absorption features being marginally resolved or under-resolved, especially at the weak end of the absorber distribution. We therefore expect the spline fit to be systematically low compared to higher resolution data and to evolve with wavelength as the density of intervening Ly$\alpha$ lines increases with redshift. To quantify and correct for this effect, we degraded our medium-resolution data to the typical resolution and S/N of the G140L data and repeated the spline fits on the degraded data. The typical ratio of the degraded-to-original spline as a function of wavelength can then serve as a correction factor for the G140L splines.  Blueward of the Galactic Ly$\alpha$ absorption feature, we assume that the correction factor is a constant, $C=0.970\pm 0.002$.  Redward of the Galactic Ly$\alpha$ absorption feature, we parameterize the correction function by fitting the median correction factor at each wavelength to a power law, $C=\left( 0.961\pm 0.001\right) \left( \frac{\lambda}{1215.67}-1\right)^{-\left(0.0045\pm 0.0005\right)}$ over roughly $\unit{1216}{\angstrom}<\lambda<\unit{1800}{\angstrom}$. We match these two functions near Galactic Ly$\alpha$ by requiring continuity. This small (2\% to 5\%) effect has little impact on the broader conclusions of this paper and suggests that our G140L spline fits are robust for the purposes of determining the large-scale spectral properties of AGNs. 

In addition to correcting for the effects of IGM and ISM absorption, the spline fits almost always interpolate successfully over absorbers intrinsic to the background AGN. In particular, at least five of our medium-resolution targets that we have retained from \citetalias{ste14} (PG 1206+459, PG 1228+416, LBQS 1435+0134, Q0232-042, PG 1630+377) have known intrinsic absorbers \citep{muz13, muz15}. These absorbers are well-resolved in the COS/G130M and COS/G160M data and are narrower than the broad undulations in the local AGN continuum, so we do not expect that they introduce any substantial error to the quantities discussed in this study. However, two of the new COS/G140L targets (PG 1115+080A1 and HE 1120+0154) feature more severe and more poorly resolved intrinsic absorption. These absorbed regions required additional manual adjustment to ensure that the semi-automated spline fitting was adequately rejecting the absorbed regions of the spectra. Though it appears that the spline fits  adequately capture the behavior of the underlying AGN continua, these two targets may suffer additional systematic error that is not present in the rest of our sample. We further discuss the potential effect of such error on our measured power-law indices in Section~\ref{sec:discslope}.

Prominent geocoronal emission lines of \HI, \ion{O}{1}, and \ion{N}{1} were masked from the spectra. We excluded 2~\AA\ on either side of \ion{N}{1} $\lambda 1200$ in both the flux array and the spline fit. We mask 9~\AA\ in the G140L flux arrays and 5~\AA\ in the G130M/G160M flux arrays on either side of \ion{O}{1} $\lambda 1304$, but we allow the spline interpolation through this region to contribute to the composite. We also mask 14~\AA\ around the Galactic Ly$\alpha$ absorption feature.  In the G140L data, all observed wavelengths greater than 2000~\AA\ were masked because the sensitivity of the COS detector declines rapidly in this region, leading to low S/N and complicated systematic effects. Finally, we discard any bins with ${\rm S/N}<0.5$, which in practice extends the long-wavelength masking slightly below 2000~\AA\ in some targets.

Each spectrum and its associated spline fit were corrected for Galactic extinction assuming a \citet{fit99} reddening law with values derived from the \citet{sch11} recalibration of the \citet{sch98} dust maps. We follow \citet{sch98} and adopt $1\sigma$ uncertainties of 16\% in the selective extinction, $E(B-V)$. We assume a mean value $R_V=3.13$ for the ratio of total-to-selective extinction and adopt a $1\sigma$ uncertainty of 0.52, following the distribution of Galactic sightlines from \citet{gem04}. We do not attempt to measure or correct for extinction intrinsic to the AGN.

Where possible, we correct the spectra for continuum \HI\ absorption due to Lyman limit systems (LLSs; $\log N_{\rm HI}\geq 17.2$) and partial-LLSs (pLLSs;  $15.0\leq \log N_{\rm HI}< 17.2$). We do not attempt to identify or correct \HI\ continuum absorption for systems with $\log N_{\rm HI}<15.0$ because these systems have a negligible effect on the spectrum, as discussed in \citetalias{ste14}. In the G140L targets, we identify these absorption systems by calculating the effective equivalent width (EW) at each pixel location by summing the contributions of spline-normalized surrounding pixels weighted by the convolution of the COS/G140L linespread function and a Voigt profile with a Doppler parameter $b=20~{\rm km~s^{-1}}$. This yields an EW array as a function of wavelength, which we search for correlated EWs at the separations of the Lyman series lines at a significance level greater than $2\sigma$ per line. We confirm these candidate \HI\ systems visually. In the medium-resolution data, we adopt the list of \HI\ systems reported in \citetalias{ste14}, although we make new column density determinations from the new coadditions of the data. We measure the EWs of the individual absorption lines and fit them to a curve of growth to determine the column density of the system and the associated uncertainty. The column density measurement is then used to correct for the \HI\ opacity shortward of the Lyman limit. We mask the region immediately surrounding the Lyman break, discarding absorber rest-frame wavelengths \unit{911.253}{\angstrom} to \unit{916.249}{\angstrom} for absorbers with $\log N_{\rm HI}<15.9$ and  \unit{911.253}{\angstrom} to \unit{920.963}{\angstrom}  for absorbers with $\log N_{\rm HI}\ge15.9$. We mask all flux at $\lambda<\unit{1161}{\angstrom}$ in the observed frame for the target HE1120+0154 because a particularly strong LLS at $z=0.273$ renders the S/N in this region too low for continuum recovery.

Five targets (PG1206+459, LBQS1435-0134, PG1522+101, Q0232-042, PG1630+377) have also been observed in the near-UV (NUV) with STIS/E230M, allowing the identification via Ly$\alpha$ (and occasionally Ly$\beta$) of \HI\ systems that are redshifted out of the COS band. For these systems, column densities were determined directly from Voigt profile fitting to the Ly$\alpha$ feature in the coadded data. These column density determinations are more uncertain than those measured in the COS data because Ly$\alpha$ is on the flat part of the curve of growth at these column densities and higher order Lyman lines are unavailable.\footnote{The one exception is the system at $z=0.9532$ in the PG1630+377 sightline, for which we were able to simultaneously fit the STIS/E230M Ly$\alpha$ feature and higher-order lines at the longest-wavelength end of the COS/G160M data. Though the COS data are noisy and contaminated by other absorption lines, they constrain the fit to a much lower value ($\log N_{\rm HI}=16.10\pm0.15$) than previously reported in the literature. \citet{jan06,jan06cat} previously reported $\log N_{\rm HI}=18.484\pm1.483$ for this system.}   The 58 identified LLS and pLLS absorption systems are listed in Table~\ref{tab:lls}. 
%final table has been inserted%
\begin{deluxetable}{lrr}
\tabletypesize{\tiny}
\tablecaption{Identified pLLS (56) and LLS (2) Absorption Systems\label{tab:lls}}
\tablecolumns{3}
\tablewidth{0pt}
\tablehead{
    \colhead{Target} & \colhead{$z_{abs}$} & \colhead{$\log N_{\rm HI}({\rm cm^{-2})}$}  \
}
\startdata
                   SDSSJ084349.49+411741.6 & 0.5326 & $16.71\pm0.08$ \\
                   SDSSJ084349.49+411741.6 & 0.5335 & $16.38\pm0.07$ \\
                   SDSSJ084349.49+411741.6 & 0.5411 & $15.47\pm0.04$ \\
                   SDSSJ084349.49+411741.6 & 0.5437 & $15.61\pm0.07$ \\
                               HE0439-5254 & 0.3280 & $15.70\pm0.02$ \\
                               HE0439-5254 & 0.6152 & $16.31\pm0.02$ \\
                               HE0439-5254 & 0.8653 & $15.57\pm0.07$ \\
                   SDSSJ100535.24+013445.7 & 0.4185 & $17.00\pm0.10$ \\
                   SDSSJ100535.24+013445.7 & 0.4197 & $15.71\pm0.03$ \\
                   SDSSJ100535.24+013445.7 & 0.8372 & $16.90\pm0.04$ \\
                   SDSSJ100535.24+013445.7 & 0.8394 & $16.28\pm0.04$ \\
                                PG1206+459 & 0.4081 & $15.75\pm0.03$ \\
                                PG1206+459 & 0.4141 & $15.47\pm0.04$ \\
                                PG1206+459 & 0.9277 & $17.20\pm0.10$ \\
               PG1206+459\tablenotemark{a} & 1.1387 & $16.42\pm0.99$ \\
                                PG1338+416 & 0.3488 & $16.46\pm0.02$ \\
                                PG1338+416 & 0.4636 & $15.37\pm0.03$ \\
                                PG1338+416 & 0.6211 & $16.17\pm0.02$ \\
                                PG1338+416 & 0.6862 & $16.49\pm0.04$ \\
                             LBQS1435-0134 & 0.2990 & $15.33\pm0.02$ \\
                             LBQS1435-0134 & 0.6119 & $15.00\pm0.04$ \\
                             LBQS1435-0134 & 0.6128 & $15.34\pm0.02$ \\
                             LBQS1435-0134 & 0.6812 & $15.52\pm0.02$ \\
                                PG1522+101 & 0.5184 & $16.24\pm0.03$ \\
                                PG1522+101 & 0.5719 & $15.85\pm0.02$ \\
                                PG1522+101 & 0.6750 & $15.88\pm0.02$ \\
                                PG1522+101 & 0.7284 & $16.60\pm0.10$ \\
               PG1522+101\tablenotemark{a} & 0.9448 & $15.25\pm0.21$ \\
               PG1522+101\tablenotemark{a} & 1.0447 & $15.59\pm0.27$ \\
               PG1522+101\tablenotemark{a} & 1.1659 & $15.41\pm0.17$ \\
                                 Q0232-042 & 0.3225 & $16.14\pm0.10$ \\
                                 Q0232-042 & 0.7390 & $16.75\pm0.10$ \\
                                 Q0232-042 & 0.8078 & $15.74\pm0.04$ \\
                Q0232-042\tablenotemark{a} & 0.9632 & $15.60\pm1.25$ \\
                Q0232-042\tablenotemark{a} & 1.0888 & $15.55\pm1.45$ \\
              HE1120+0154 & 0.2558 & $       >18.0$ \\
                               HE1120+0154 & 0.5771 & $15.71\pm0.13$ \\
                                PG1630+377 & 0.2741 & $17.05\pm0.10$ \\
                                PG1630+377 & 0.2782 & $15.09\pm0.05$ \\
                                PG1630+377 & 0.4177 & $15.71\pm0.02$ \\
                                PG1630+377 & 0.8110 & $15.64\pm0.03$ \\
                                PG1630+377 & 0.9145 & $15.94\pm0.06$ \\
               PG1630+377\tablenotemark{a} & 0.9532 & $16.10\pm0.15$ \\
               PG1630+377\tablenotemark{a} & 0.9596 & $15.77\pm0.30$ \\
               PG1630+377\tablenotemark{a} & 1.0961 & $16.12\pm0.58$ \\
               PG1630+377\tablenotemark{a} & 1.1608 & $15.30\pm0.41$ \\
               PG1630+377\tablenotemark{a} & 1.4333 & $15.27\pm0.23$ \\
                               HE0248-3628 & 0.5346 & $16.61\pm0.10$ \\
                               HE0248-3628 & 0.5443 & $15.65\pm0.04$ \\
                               HE0248-3628 & 0.5704 & $15.43\pm0.06$ \\
                                    US2504 & 0.3498 & $15.58\pm0.13$ \\
                   SDSSJ125140.83+080718.4 & 0.4412 & $15.60\pm0.16$ \\
                   SDSSJ125140.83+080718.4 & 0.7329 & $16.03\pm0.14$ \\
                   SDSSJ094209.14+520714.5 & 0.7411 & $16.25\pm0.18$ \\
                              PG1115+080A1 & 0.5964 & $15.27\pm0.16$ \\
                               SBS1307+462 & 0.3679 & $15.45\pm0.30$ \\
                               SBS1307+462 & 0.4601 & $15.52\pm0.27$ \\
                               SBS1307+462 & 0.5176 & $16.31\pm0.53$ \\
\enddata
\tablenotetext{a}{Measurement made using STIS/E230M data.}
\end{deluxetable}

We characterize the local continuum of each spectrum as a power law,
\begin{equation}
F_\lambda = F_0\left( \frac{\lambda}{\rm 1100~\AA}\right)^{\alpha_\lambda},
\end{equation}
where the observed flux distribution, $F_\lambda$, is a function of the rest-frame wavelength, $\lambda$ and has two free parameters: the normalization, $F_0$, and the power-law index, $\alpha_\lambda$. The index can be converted to the equivalent flux distribution in frequency space, $F_\nu\propto \nu^{\alpha_\nu}$, via $\alpha_\nu=-(2+\alpha_\lambda)$. The best-fitting power law passes through two spline points such that all other possible power laws through other spline-point combinations lie above the best-fit power law. We determine the probability distribution of the best-fit power law via 15,000 Monte Carlo realizations of the spectrum-correction process that account for the aforementioned uncertainties. In each realization, the parameters are calculated iteratively and analytically, enforcing the additional constraints that the two points must be separated by at least 50~\AA\ and lie more than 5~\AA\ from the edge of the spectrum to avoid effects from the spline ends. The best-fit parameters and 1$\sigma$ marginalized uncertainties are determined from this distribution. These uncertainties are dominated by the extinction correction and LLS/pLLS measurements. We emphasize that this parameterization represents only the local behavior of the spectrum, and it may not represent exclusively continuum flux if the broad emission lines overlap such that there are no  pure-continuum windows in the spectrum. We also note that these power-law fits do not account for the possibility of unidentified \HI\ continuum absorption from LLSs whose Lyman breaks are redshifted out of the COS observational bands (approximately $\lambda_{\rm obs}>\unit{2000}{\angstrom}$ for G140L and $\lambda_{\rm obs}>\unit{1800}{\angstrom}$ for G160M) but are not observed with any NUV data. Though we correct for these effects statistically in our composite spectrum, such corrections to individual objects are not particularly useful, owing to the stochasticity of such \HI\ systems. As a result, each of these power-law slopes should be treated as a lower limit to the true slope until near-UV observations of these objects become available. The best-fit power-law parameters are given in Table~\ref{tab:targs}.

\begin{figure*}
%final figure has been inserted%
\epsscale{0.90}
\plotone{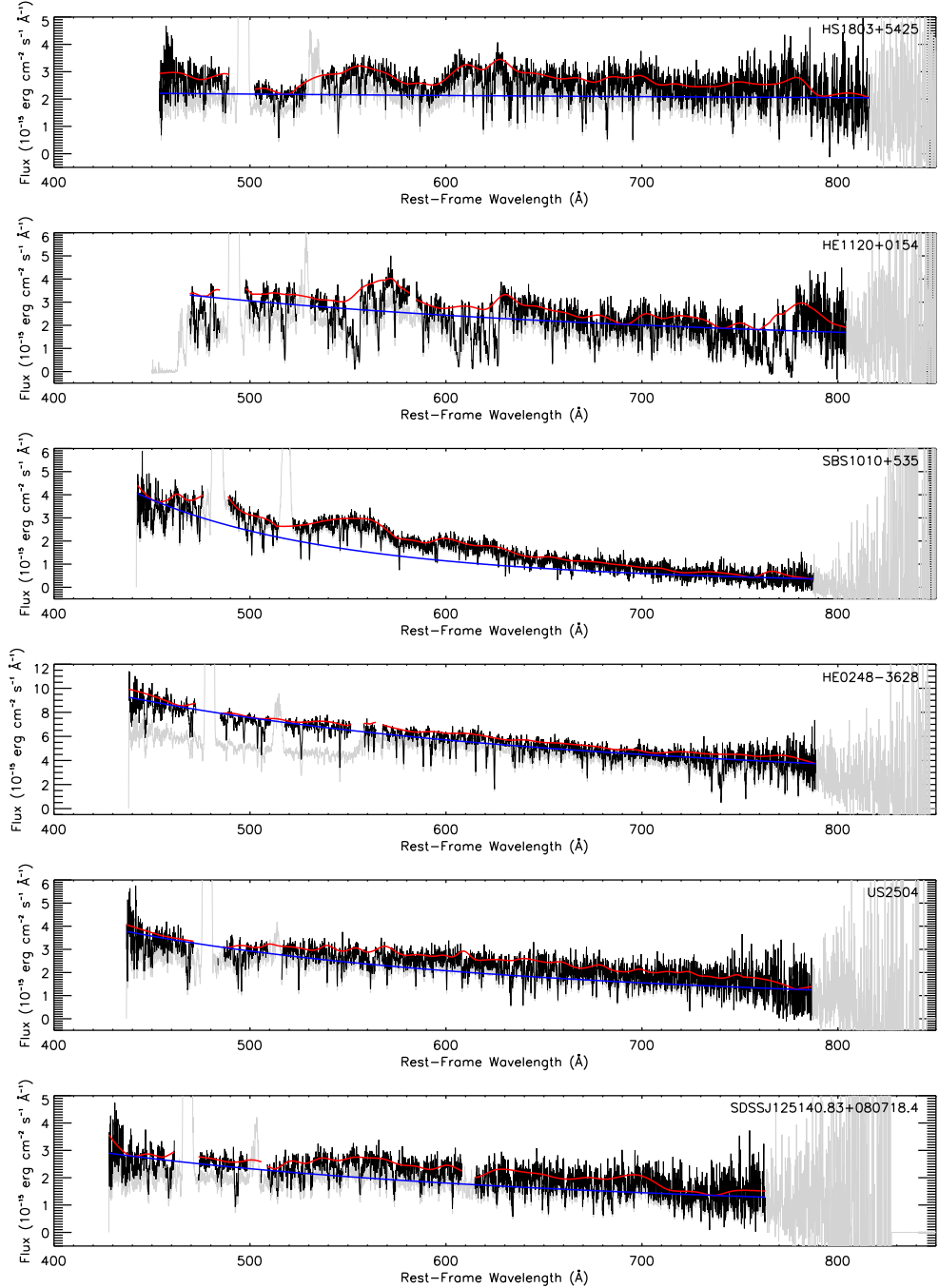}
\caption{AGN spectra observed with COS/G140L. The light-gray line is the uncorrected flux, while the black line is  corrected for Galactic extinction, intervening \HI\ absorption, and geocoronal emission as described in Section~\ref{sec:targmeth}. The red line is the spline fit, and the blue line is a power-law fit to the underlying continuum.  \label{fig:specs1}}
\end{figure*}

\begin{figure*}
%final figure has been inserted%
\epsscale{1.0}
\plotone{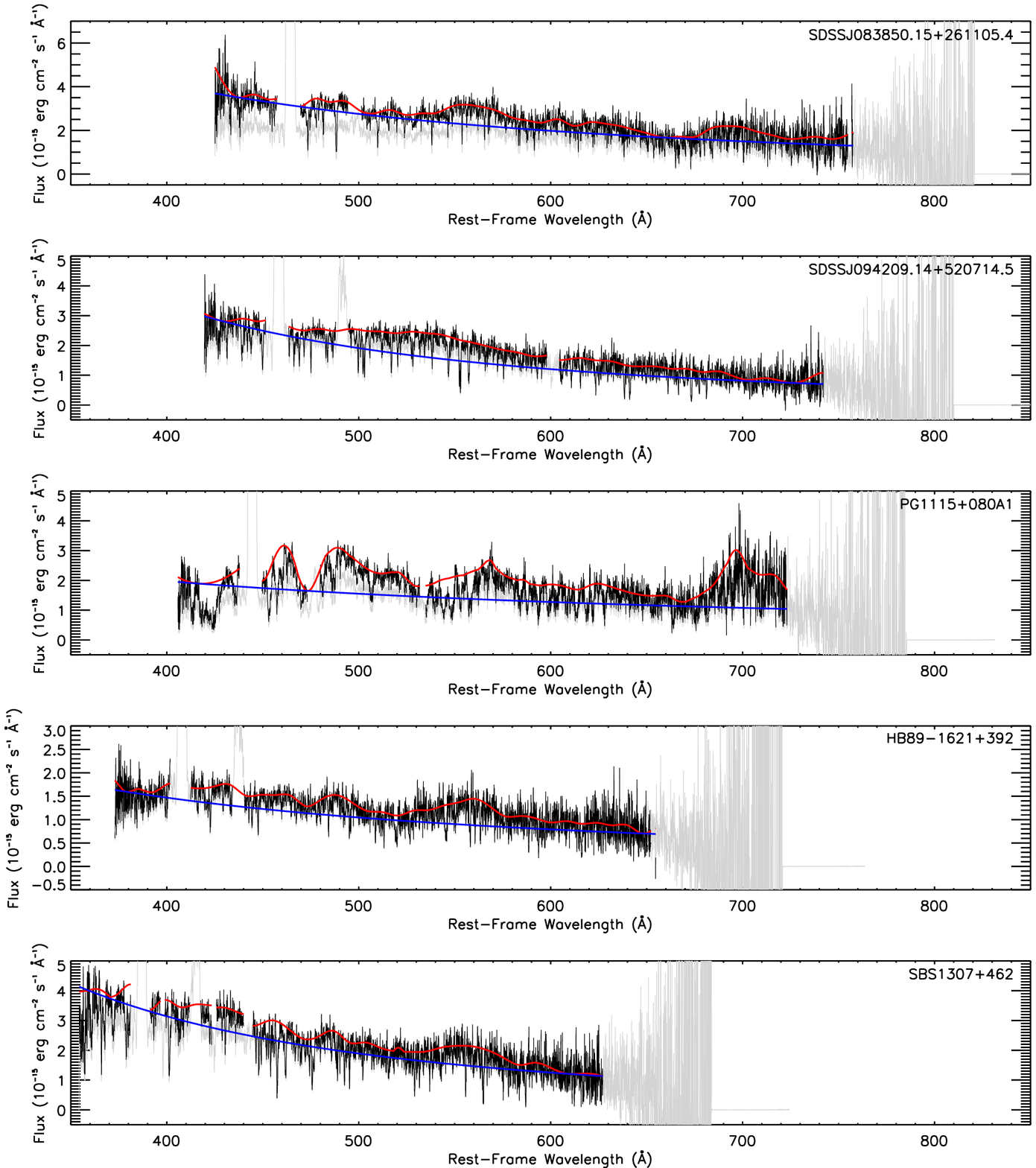}
\caption{Same as Figure~\ref{fig:specs1}.\label{fig:specs2}}
\end{figure*}

The eleven G140L spectra  are plotted in Figures~\ref{fig:specs1} and \ref{fig:specs2}. The light-gray line is the observed flux, and the overplotted black line is the flux corrected for Galactic extinction, \HI\ absorption, and geocoronal emission. This corrected flux is used to construct our composite spectrum. The solid red line is the associated spline fit corrected for the measured resolution and wavelength-dependent systematic error, while the solid blue line is a power-law fit to the continuum underlying the spline fit. 

\subsection{Composite Construction}
\label{sec:compconst}
In order to construct a composite spectrum, the spectra and their associated spline fits must first be resampled onto uniform wavelength grids. As in \citetalias{shu12a} and \citetalias{ste14}, we adopt the method formalized in Equations 2 and 3 of  \citet{tel02} and resample onto 0.61~\AA\ bins. Because we are focused on a more limited wavelength regime in this study compared to \citetalias{ste14}, the composite construction process can be simplified. Instead of the bootstrap-normalization process, we normalize  the resampled spectra to their median  flux values in a single window of overlap, $575~{\rm \AA}<\lambda <610~{\rm \AA}$, without any apparent strong emission lines. We construct two styles of composite spectra. The  first style is calculated as the geometric mean of the contributing flux bins and is used for slope determinations because it preserves power-law shapes. The second style uses the median of the contributing flux bins to better preserve the relative strengths of the emission lines.

In both cases, in a method analogous to that used for the individual target spectra, we construct 15,000 Monte Carlo realizations of mock composites to determine the probability distributions of measured quantities. Unlike the realizations  used for the determination of power-law parameters for the individual targets, we  statistically account for the effects of unidentified LLSs and pLLSs whose \unit{912}{\angstrom} Lyman edges are redshifted out of the observational band. To do so, however, we must adopt a functional form for the evolution of the distribution of absorbers in both column density and redshift. Following the usual notation in the literature, we parameterize this frequency distribution as a power law,
\begin{equation}
f(N_{\rm HI}, z)=A\left(\frac{N_{\rm HI}}{{\rm cm^{-2}}}\right)^{-\beta}(1+z)^{\gamma},
\end{equation}
over the ranges $10^{15}<N_{\rm HI}<10^{18}~{\rm cm^{-2}}$ and $1.19<z<2.14$. Mock absorbers are drawn from this distribution according to Poisson statistics over the unobserved redshift range in which they could have occurred. The modeled effects of their continuum opacity are then injected into each mock composite realization.

Such an absorber distribution is well-characterized at both low-redshifts \citep[e.g.,][]{til12,dan14} and high-redshifts \citep[e.g.,][]{rud13} where extensive FUV and optical observations probe the Ly$\alpha$ forest. However, results for the intermediate redshifts of relevance here, best studied in the NUV, are substantially more uncertain. The few studies that have been conducted neither arrive at consistent results nor conduct thorough, multivariate analyses, making the optimal choices for $(A, \beta, \gamma)$ unclear. Further, no one study has covered the entirety of the relevant range of both column densities and redshifts, forcing us to attempt to match disparate distributions of absorbers from different methodologies. Our general approach is to choose values for these parameters based on measurements from the literature and allow them to vary in the Monte-Carlo realizations of the composite spectrum to account for their uncertainties.

The slope of the column-density distribution is perhaps the most secure of the three parameters, as it remains fairly constant near $\beta\approx 1.6$ over a wide variety of redshifts and column density ranges \citep{kim02,jan06,til12,rud13,dan14}. We therefore adopt $\beta=1.6\pm 0.1$ in our absorber distribution. The overall normalization of the distribution and its redshift evolution remain far more uncertain. Early surveys \citep{sto94,ste95} used combinations of \textit{International Ultraviolet Explorer} and \textit{HST} data to characterize the LLS absorber distribution as a function of redshift alone, obtaining $\gamma=1.5\pm 0.39$ for $\log N_{\rm HI}\ge 17.2$ over $0.32\le z_{\rm LLS}\le 4.11$. More recently, \citet{kim02} combined ground-based data and \textit{HST} surveys to obtain $\gamma=3.11\pm 0.42$ for $14.5<\log N_{\rm HI}<17$ over $1.5<z_{\rm abs}<4$. We attempt to accommodate the range of results in the literature by adopting $\gamma=2\pm 1$. The overall normalization for this distribution is constrained by the expectation that the density of absorbers should increase with redshift; the low-redshift and high-redshift surveys thus bound the normalization. We therefore set the value of $A$ by requiring that the integrated quantity $f(N_{\rm HI}, z)=5.3\pm 0.25$ over $10^{15}<N_{\rm HI}<10^{18}~{\rm cm^{-2}}$ and $1.19<z<2.14$, in approximate agreement with the \citet{ste95} and \citet{kim02} integrated distributions and the bounds set by \citet{dan14} and \citet{rud13}. Though none of these studies investigates the joint-probability distribution for these three parameters, the normalizations of their fits to the redshift distribution of absorbers imply an additional constraint on the integrated column-density distribution, which we implement by requiring that $0.2<A\int_{10^{15}}^{10^{18}}N_{\rm HI}^{-\beta}dN_{\rm HI}<9$. Finally, we require that no realization yields an \HI\ opacity that implies an intrinsic AGN spectral slope of $\alpha_\nu>+1.15$, roughly $\Delta\alpha_\nu=0.9$ greater than the maximum slope we observed at lower redshifts in \citetalias{ste14}.

The Monte Carlo experiment implements a bootstrap-with-replacement technique to quantify the effects of cosmic variance among our targets. Each realization is composed of a randomly selected set of available targets. Because of the small number of spectra contributing at some wavelengths, a few combinations lead to composite spectra that lack data at wavelengths of interest; these realizations are rejected. The final composite spectrum is determined from the median flux in each wavelength bin of the Monte Carlo realizations. Measured quantities such as the power-law parameters are determined directly from the ensemble of realizations in the same manner as for the individual targets.

\begin{figure*}
%final figure has been inserted%
\epsscale{1.15}
\plotone{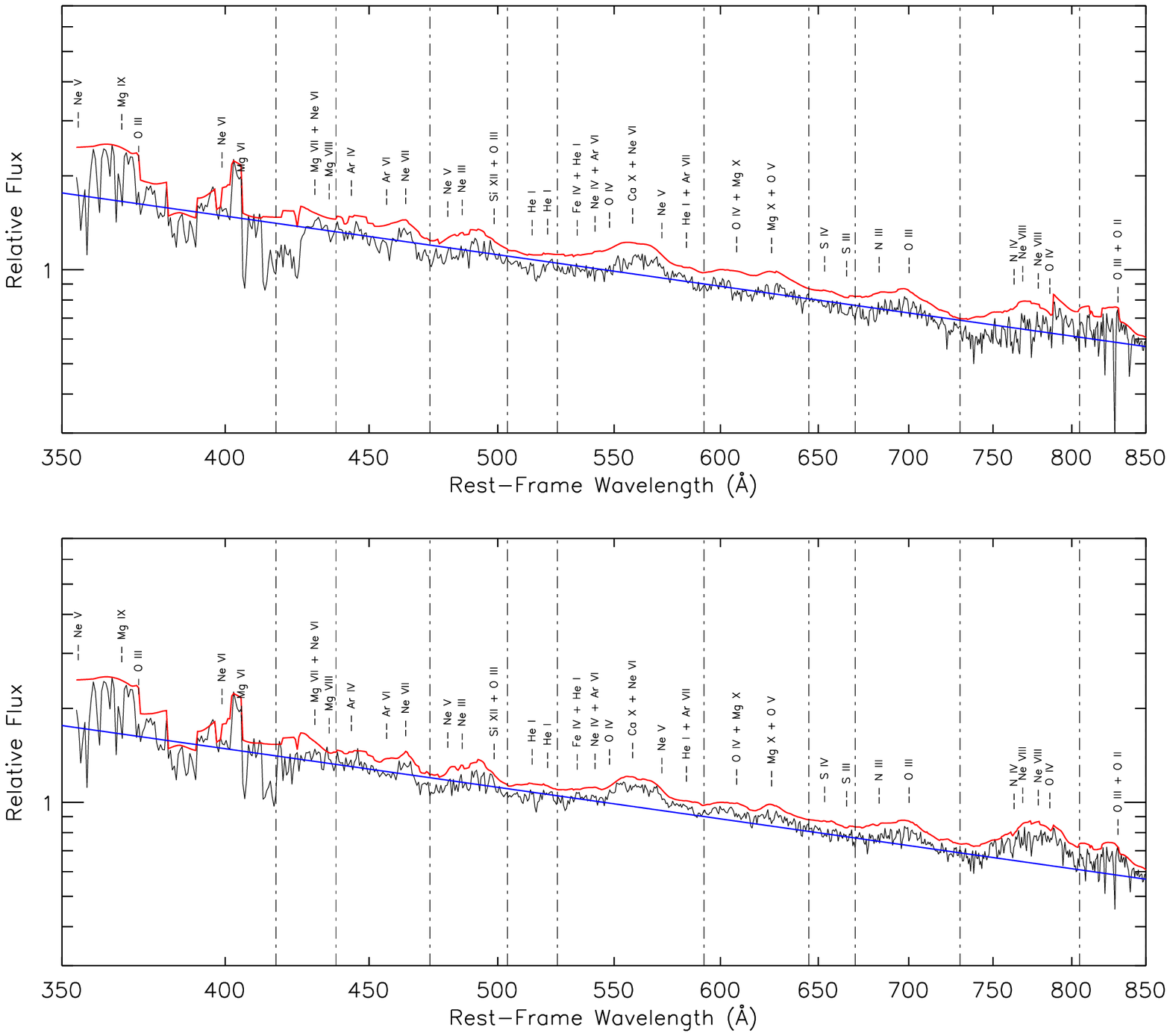}
\caption{Composite spectra for the full sample of 20 AGN. The top panel is the geometric-mean composite, and the bottom panel is the median composite. Black lines show the composite flux, and red lines show the composite spline, which lies above the black line owing to intervening absorption. Possible emission-line identifications are labeled. The best-fit continuum power law ($\alpha_\nu=-0.72\pm0.26$) is shown in blue. Vertical dot-dashed lines indicate the divisions used for emission-line EW measurements in Section~\ref{sec:emission}.\label{fig:comp}}
\end{figure*}

\begin{figure}
%final figure has been inserted%
\epsscale{1.1}
\plotone{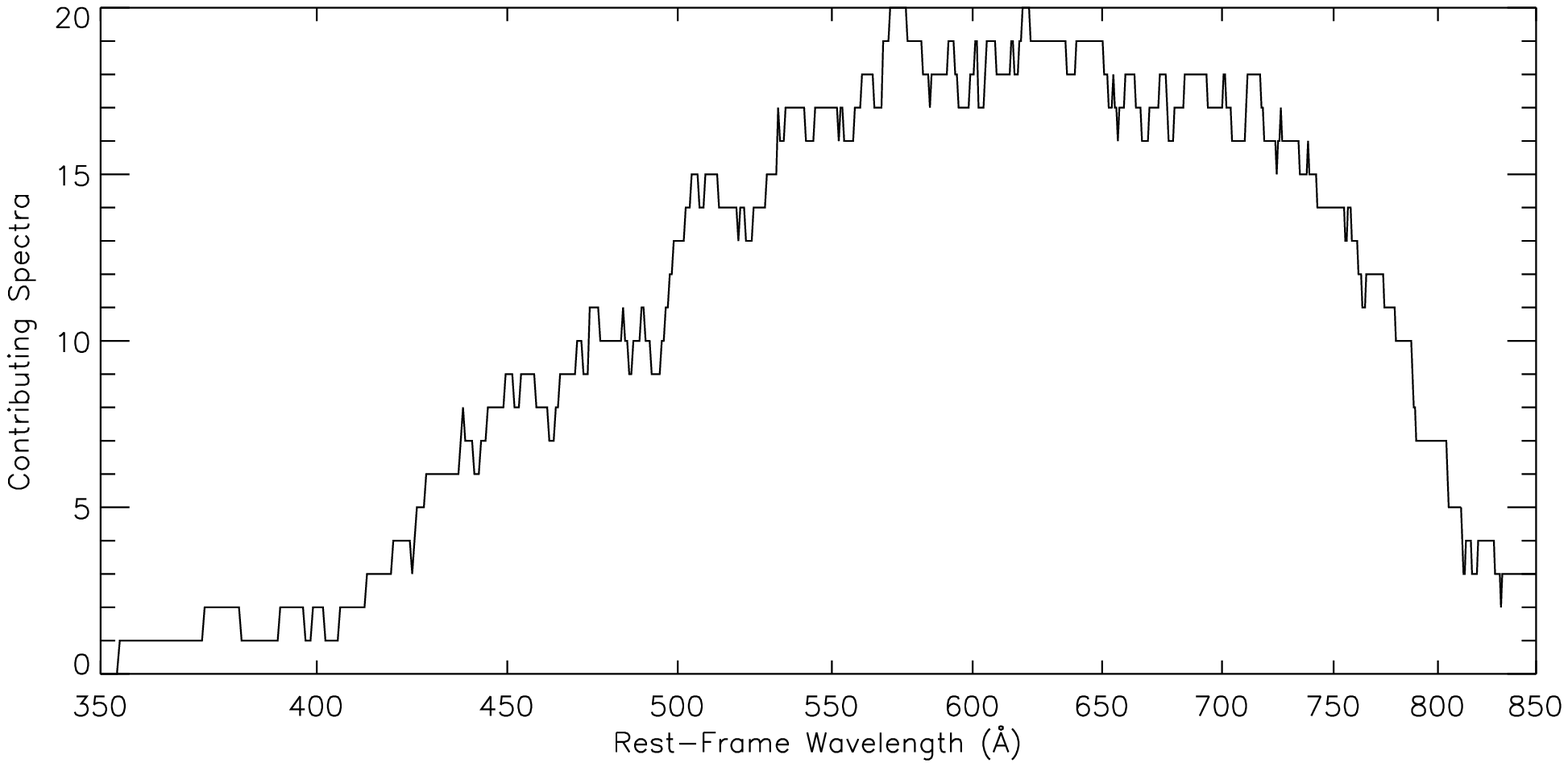}
\caption{The number of AGN spectra contributing to the composite spectra as a function of rest-frame wavelength.\label{fig:numcontrib}}
\end{figure}

Our full-sample composite (geometric-mean and median) spectra are presented in Figure~\ref{fig:comp} along with the best-fit power law, $\alpha_\nu =-0.72\pm 0.26$, as determined from the Monte-Carlo realizations. The solid red line is the composite spline spectrum that best represents the intrinsic AGN spectrum shape, while the black line is the composite flux spectrum, produced directly from the corrected flux of the individual targets. Potential identifications of the major emission lines are labeled in the plot. Figure~\ref{fig:numcontrib} shows the number of AGNs contributing to the composite spectra as a function of wavelength. Because the number of contributing spectra fall off rapidly below 450~\AA\ and above 770~\AA, we restrict all power-law measurements to this wavelength range. The long wavelength fall off is due to our sample selection and results from our redshift threshold that allows all targets to be normalized to the same continuum window ($575~{\rm \AA}<\lambda <610~{\rm \AA}$). The short wavelength fall off, on the other hand, represents a true paucity of COS observations of AGNs that probe this wavelength regime. 

\section{Discussion}
\label{sec:disc}
\subsection{The EUV Slope from 450 to 770~\AA}
\label{sec:discslope}

One of the primary goals of this study is the determination of the typical slope of AGN spectra in the EUV wavelength regime for input to radiative transfer calculations of quantities such as the UVB or photoionization models of the IGM and gas in and around active galaxies. As mentioned in Section~\ref{sec:compconst}, our full-sample composite spectrum leads to a power-law index of  $\alpha_\nu = -0.72\pm 0.26$. This value is substantially harder than EUV spectral slopes reported from most other composites, including \citetalias{shu12a} ($\alpha_\nu=-1.41 \pm 0.20$), \citetalias{ste14} ($\alpha_\nu=-1.41 \pm 0.15$), \citet{tel02} with $\alpha_\nu=-1.57\pm 0.17$ in the radio-quiet subsample, and \citet{lus15} with $\alpha_\nu=-1.70 \pm 0.61$. The harder \citet{sco04} slope, $\alpha_\nu=-0.56^{+0.38}_{-0.28}$, is in better agreement with this study's result, although their \textit{FUSE} composite only extends down to \unit{630}{\angstrom}. In \citetalias{ste14}, we explained the discrepant \citet{sco04} slope as a result of their choice of continuum windows and their fitting the power law over the tops of unidentified emission lines in their shortest-wavelength fitting window ($630-750~{\rm \AA}$). This resulted in an artificial hardening of the spectrum and incorrect separation of the power-law continuum and the broad emission-line flux.

The apparent tension between our present result and the slopes of other composites can be explained several ways. It is important to note that, while all of these composite spectra probe the EUV, their wavelength coverage is not completely equivalent, and the windows used to determine the power-law slope differ among surveys. For example, the \citet{tel02} EUV power law is a fit to many flux bins that range from 350~\AA\ to 1150~\AA\ with varying degrees of contamination from broad emission lines, but it is dominated by the higher-S/N bins at the longer-wavelength end of this range. The authors noted that the spectral slope hardened substantially below $\sim 500~\AA$, but they ascribed little significance to the change, owing to the small number of spectra contributing at those wavelengths.  \citet{sco04} used the same fitting windows as \citet{tel02}, but enhanced emission-line flux in some of their targets may have led to a harder slope, as we discussed in \citetalias{ste14}. \citet{lus15} adopt the fitting windows initially defined by \citet{tel02}, but they note that the fit is complicated by the the high density of emission lines in the region, rendering the power-law model an insufficient description of the spectrum. They explain their softer slope compared to \citet{sco04} in terms of improvements to their statistical IGM absorption corrections. Finally, our \citetalias{ste14} power law is determined by connecting two continuum-like windows centered on 724.5~\AA\ and 859~\AA, chosen to avoid the locations of expected strong emission lines. Therefore, it does not constrain the slope at wavelengths below $\sim 724~{\rm \AA}$, a region that can be seen to harden in their Figure~5, much like the \citet{tel02} composite.

The power law described by our current study therefore probes a  different spectral region than in \citetalias{ste14}. This region ($450~{\rm\AA}<\lambda<770~{\rm\AA}$)  exhibits a spectral hardening seen in several previous studies, which also might be expected from the need for the spectrum to match up with soft X-ray observations with $F_{\nu}\propto \nu^{-1}$ \citep[e.g.,][]{has96}. The power-law determination in this study further differs from the others because it does not specify any continuum windows \textit{a priori} and thus avoids the need to explain the unidentified and seemingly unphysical ``absorption'' features that arise in the \citet{tel02}, \citet{sco04}, and \citet{lus15} composites.

\begin{figure}
%final figure has been inserted%
\epsscale{1.15}
\plotone{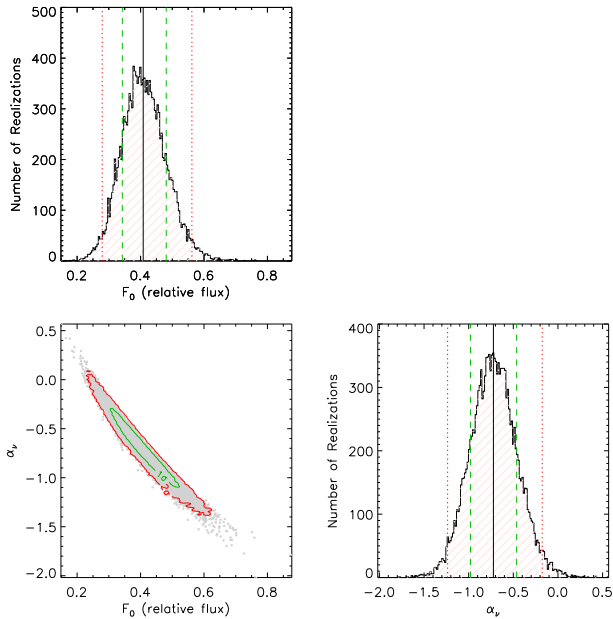}
\caption{The results of the Monte-Carlo realizations for the power-law parameters. The lower-left panel plots the 2D joint-probability distribution for the power-law index, $\alpha_\nu$, and the normalization, $F_0$, with approximate $1\sigma$ and $2\sigma$ contours in green and red, respectively. The other panels show the marginalized distributions for these parameters, with the median value ($\alpha_\nu =-0.72\pm 0.26$) marked with a solid black line and the $1\sigma$ and $2\sigma$ bounds marked with green dashed lines and red dotted lines, respectively. \label{fig:2ddist}}
\end{figure}

It might also be possible to explain the hardening of the spectral slope through errors in our corrections for Galactic extinction or \HI\ absorption, either from the measured column densities or the statistical correction for unidentified systems. Our Monte Carlo experiments, however, attempt to account for these uncertainties. The two-dimensional joint-probability distribution for the full suite of power-law realizations is shown in Figure~\ref{fig:2ddist}, along with the marginalized distributions for $F_0$ and $\alpha_\nu$. Though these correlated variables show a low-probability tail toward softer spectra caused by the unidentified \HI\ systems, a steady or softening slope compared to the EUV slope of \citetalias{ste14} is excluded at $99.6$\% confidence.  The hardening of the EUV slope observed in our composite is too large to be explained by uncertainties in Galactic extinction or \HI\ absorption alone.

As mentioned in Section~\ref{sec:targmeth}, two of the targets (PG 1115+080A1 and HE 1120+0154) could potentially introduce systematic error to our measurements because of the presence of absorption associated with the AGN. The bootstrap-with-replacement procedure included in our Monte Carlo experiments naturally includes in the error analysis any additional variance carried by these two targets. However, as an additional check to ensure that the inclusion of these targets is not artificially hardening the power-law slope, we repeated the Monte Carlo calculations with these two targets removed. This procedure results in a slightly harder EUV slope, $\alpha_\nu=-0.66\pm0.29$, consistent with the full-sample results.

\begin{figure}
%final figure has been inserted%
\epsscale{1.15}
\plotone{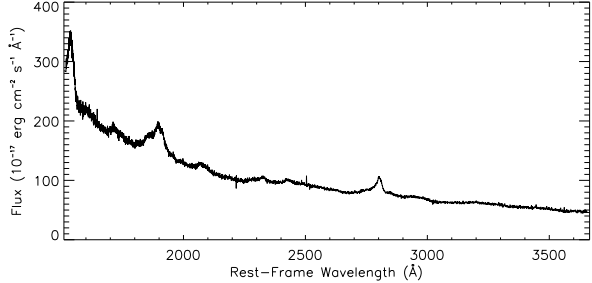}
\caption{Rest-frame FUV spectrum of SBS1010+535 ($z_{\rm AGN}=1.515848$) from the Sloan Digital Sky Survey. Note the rapidly hardening slope and the absence of any strong, intervening \ion{Mg}{2} absorbers shortward of 2800~\AA\ in the range $1.190<z<1.516$.\label{fig:sdss}}
\end{figure}

\begin{figure*}
%final figure has been inserted%
\epsscale{1.15}
\plotone{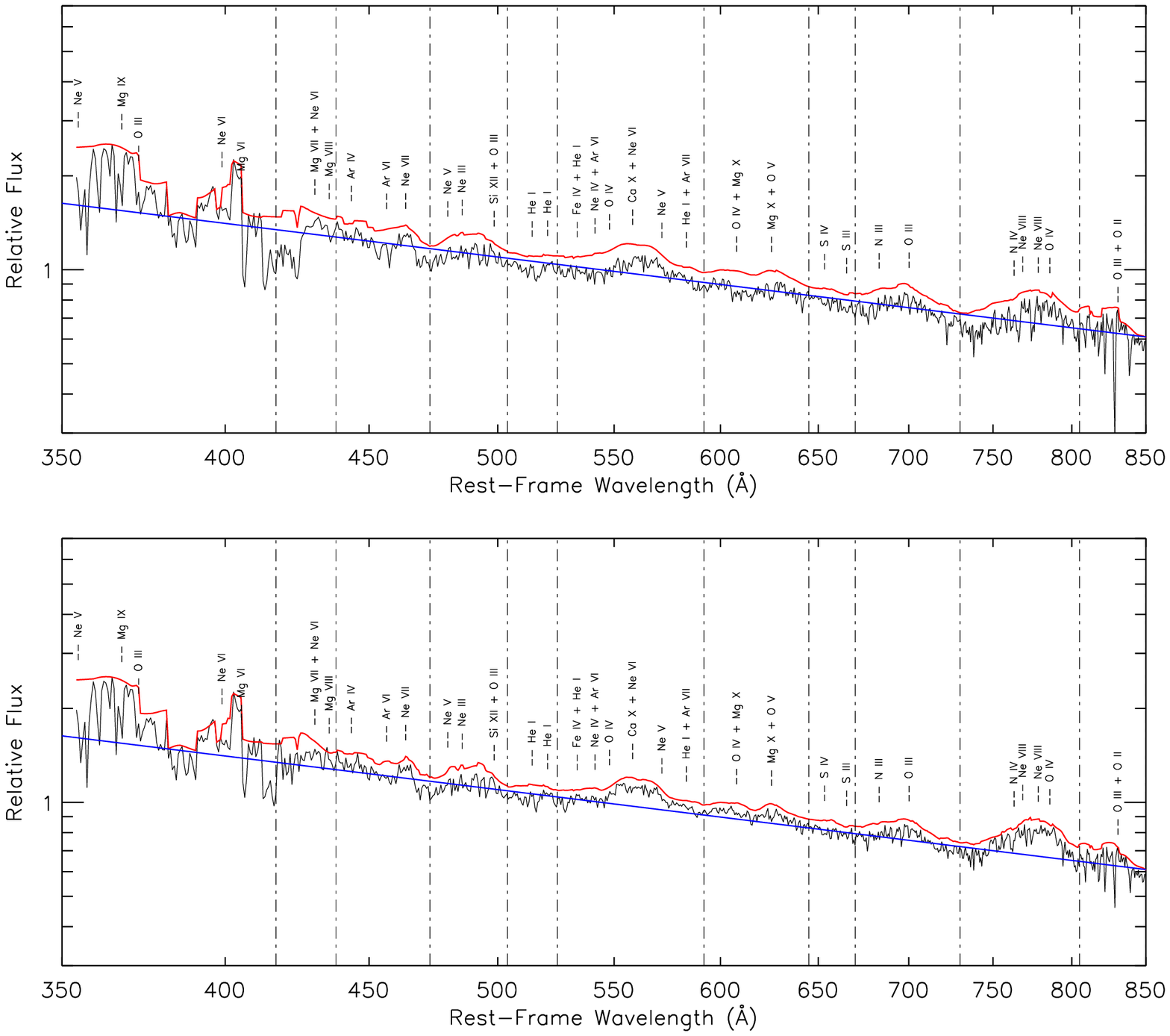}
\caption{Same as Figure~\ref{fig:comp}, but with the outlier SBS1010+535 removed from the sample. The power-law index is $\alpha_\nu = -0.89\pm 0.22$, slightly steeper than the full sample.\label{fig:nolowcomp}}
\end{figure*}

At least one of our targets, SBS1010+535, does exhibit a very hard, inverted slope, $\alpha_\nu =+2.17\pm 0.01$, that warrants further discussion. The most obvious explanation for such an anomolous slope is the potential presence of an extremely strong, undetected \HI\ system at wavelengths longward of our observational band. Because this AGN ($z_{\rm AGN}=1.515848$) has never been observed in the near-UV, we cannot directly investigate the \HI\ systems in the relevant redshift range. However, the object has publicly-available optical spectra from the Sloan Digital Sky Survey \citep[SDSS; for the most recent data release see][]{ala15} which gives some indications about the intrinsic spectral slope at longer, rest-frame FUV wavelengths and could potentially reveal the redshifts of intervening \HI\  absorption systems via the \ion{Mg}{2} $\lambda 2800$ feature. This spectrum is shown in Figure~\ref{fig:sdss}. Even in the FUV, the spectrum is anomalously hard, exceeding all but one of the slopes discussed in \citetalias{ste14}. At rest wavelengths $\lambda_{\rm rest}>2000$~\AA, the slope is $\alpha_\nu\approx -0.45$, but it rapidly hardens at shorter wavelengths to  $\alpha_\nu\approx +0.29$. No detectable (${\rm EW}>\unit{0.3}{\angstrom}$) \ion{Mg}{2} absorbers are apparent in the expected range ($\unit{6130}{\angstrom}<\lambda_{\rm obs}<\unit{7041}{\angstrom}$), suggesting that if a LLS is present, it must be part of a low-metallicity population. Previous surveys \citep{leh13} have identified a population of LLS absorbers with $\sim2.5\%$ solar metallicity. When combined with our \ion{Mg}{2} non-detection, such a metallicity would imply a maximum \HI\ column density of $\log N_{\rm HI}\sim 18.67$. Therefore, a low-metallicity LLS with sufficient column density  to account for the anomalous EUV slope may be present despite our non-detection of \ion{Mg}{2} absorbers. Further, the longest wavelength ($\sim\unit{1800}{\angstrom}$) fluxes from the COS/G140L data are a factor of $\sim 4$ lower than at the shortest wavelength ($\sim\unit{3800}{\angstrom}$) covered by the SDSS data, perhaps suggesting the effects of a LLS between the two regions (however, target variability may also contribute, as these observations were not simultaneous). The most likely explanation for the  observed inverted EUV slope of SBS1010+535  seems to be a combination of its position as one of the hardest slopes in the AGN slope distribution and additional hardening from one or more intervening \HI\ absorbers in the  range $1.190<z<1.516$. To test the effect of including this object in our sample, Figure~\ref{fig:nolowcomp} shows a version of the composite with it removed, which leads to a slope $\alpha_\nu = -0.89\pm 0.22$ using the same methodology as above. Although this softens the composite slope somewhat, it is still significantly harder than the EUV slopes from \citet{tel02} or \citetalias{ste14}. The inclusion of the outlier SBS1010+535 in our sample is thus insufficient to explain the observed spectral hardening of the composite.

If the fast-rising flux at short wavelengths is not an artifact of a systematic uncertainty associated with foreground corrections, then we must understand it in terms of flux from the AGN itself. It is plausible that the power-law slope represents a true hardening of the accretion disk continuum spectrum in the EUV. This EUV-rise has been predicted by some models, such as those that include reprocessing of the disk continuum by cold, optically thick clouds \citep{law12}. However, more standard alpha-disk and wind models \citep[e.g.,][and references therein]{slo12,lao14} tend to predict EUV spectral turnovers that depend on the innermost edge of the disk, determined by black hole mass and spin, and on the radial profile of the mass accretion rate. These sorts of models are difficult to simultaneously reconcile with our observed hardening EUV slope and with the shorter-wavelength EUV  and FUV slopes seen by \citetalias{ste14}.

Perhaps the simplest explanation for the hardening spectral index is the large number of highly blended broad emission lines in the spectral region between 400 and 800~\AA. Far more than in the optical, near-UV, or FUV, these overlapping emission lines create a pseudo-continuum, leaving no spectral region free of emission-line flux. As noted by \citet{lus15}, the simple power-law model that has often  been used  may be insufficient to determine the behavior of the underlying continuum separately from the broad emission-line flux. This pseudo-continuum limits the utility of EUV power-law fits to composite spectra for the purpose of understanding the generation of continuum flux within accretion disk models, but it is not particularly problematic from the standpoint of determining inputs to photoionization modeling or calculations of the UVB. Such calculations are concerned only with the average contribution of photons from AGN to the background at each wavelength. An ideal determination of the UVB would include all the photons,  whether they originate in the accretion disk or the broad-line region. However, if the power-law parameterization is substantially contaminated with emission-line flux, it cannot be used to extrapolate the continuum slope beyond the observational window.

\subsection{Emission Lines}
\label{sec:emission}
\begin{figure}
%final figure has been inserted%
\epsscale{1.15}
\plotone{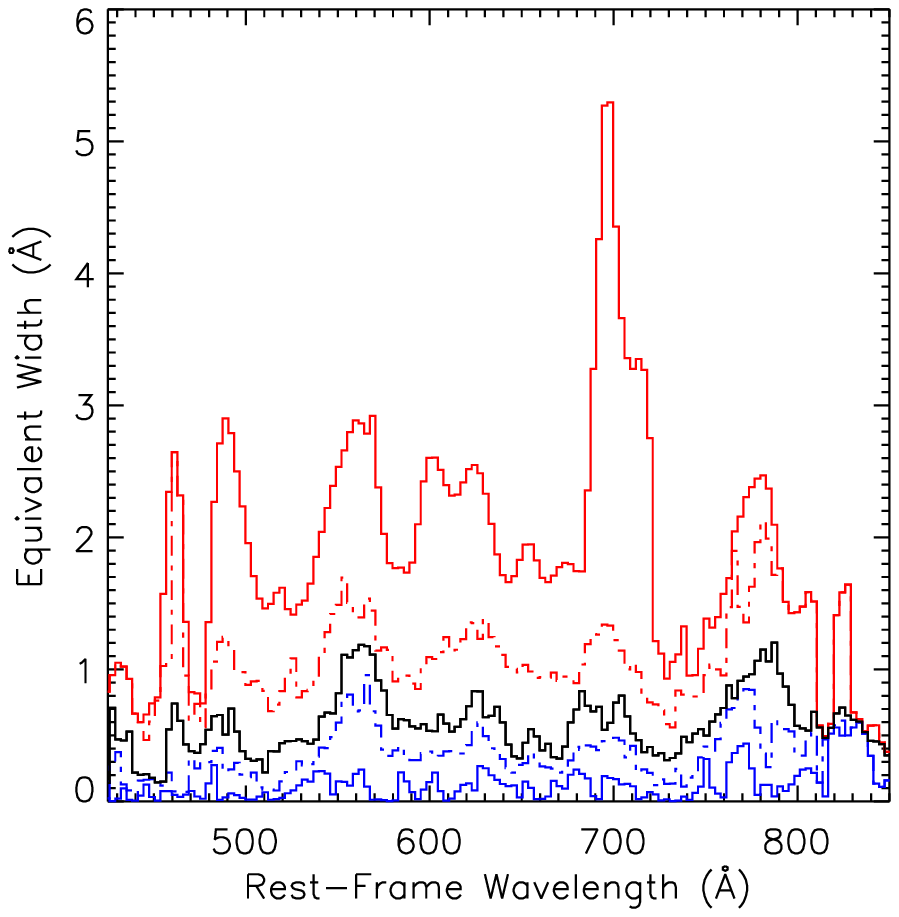}
\caption{Emission-line equivalent width for the sample of targets as a function of AGN rest-frame wavelength, illustrating the diversity of emission line strengths among these targets. The lines separate the contributing AGNs into quartiles at each wavelength, from bottom to top: bottom (blue) solid line plots the minimum values;  bottom (blue) dot-dashed line plots the 25th percentile;  middle (black) solid line plots the median values;  top (red) dot-dashed line plots the 75th percentile; and  top (red) solid  line plots the maximum values. Bins ($\sim 3~{\rm\AA}$) are five times the width of the bins used in the composite spectrum.\label{fig:ewplot}}
\end{figure}

%final table has been inserted
\begin{deluxetable*}{lcrr}
\tabletypesize{\tiny}
\tablecaption{Equivalent Widths of Spectral Regions of Emission\label{tab:ews}}
\tablecolumns{4}
\tablewidth{0pt}
\tablehead{
    \colhead{Possible Contributing Ions} & \colhead{Wavelength Range ({\AA})} & \colhead{EW ({\AA})} & \colhead{EW ({\AA})}  \\
     & &  & \colhead{(no SBS1010+535)}  \
}
\startdata
                               Mg VII, Ne VI, Mg VIII & $417-438$ & $  2.2\pm 1.1$ & $  3.5\pm 1.1$ \\
                                 Ar IV, Ar VI, Ne VII & $438-473$ & $  2.6\pm 0.9$ & $  3.5\pm 1.0$ \\
                          Ne V, Na III, Si XII, O III & $473-504$ & $  2.3\pm 0.5$ & $  3.0\pm 0.6$ \\
                                                 He I & $504-525$ & $  0.6\pm 0.4$ & $  0.9\pm 0.4$ \\
 Fe IV, He I, Ne IV, Ar VI, O IV, Ca X, Ne VI, Ar VII & $525-592$ & $  7.9\pm 0.9$ & $  8.1\pm 1.1$ \\
                                      O IV, Mg X, O V & $592-645$ & $  5.9\pm 0.8$ & $  5.2\pm 0.9$ \\
                                          S IV, S III & $645-670$ & $  2.0\pm 0.4$ & $  1.4\pm 0.4$ \\
                                         N III, O III & $670-730$ & $  7.2\pm 1.0$ & $  5.2\pm 0.9$ \\
                                  N IV, Ne VIII, O IV & $730-805$ & $ 16.4\pm 2.2$ & $ 12.8\pm 2.0$ \\
                                          O III, O II & $805-850$ & $  7.9\pm 2.5$ & $  4.3\pm 2.3$ \\
\enddata
\end{deluxetable*}

The large number and diverse strengths of the emission lines in the EUV has been apparent in all composite spectra that cover this spectral region. Unfortunately, perhaps owing to the difficulties associated with measuring the individual line fluxes, there have been few attempts to understand and disentangle this complicated ensemble of lines. Though there have been some attempts to reproduce EUV emission lines at $\lambda\geq 765~{\rm \AA}$ (e.g., \ion{N}{4} $\lambda 765$, \ion{N}{3} $\lambda 991$, \ion{O}{4} $\lambda 788$, \ion{O}{3} $\lambda 834$, \ion{O}{2} $\lambda 833$) via photoionization modeling \citep[e.g.,][]{mol14}, there has been no detailed effort to do so at shorter wavelengths. It is therefore difficult to uniquely identify the emission lines that dominate our composite spectra without a detailed photoionization study beyond the scope of this paper. Instead, we have labeled Figures~\ref{fig:comp} and \ref{fig:nolowcomp} with  possible line identifications. These identifications are based on the strongest lines in each spectral region as estimated from the \textsc{Cloudy} \citep[last described by][]{fer13} models used by \citet{mol14} using their default ``local optimally emitting cloud'' model \citep[see][]{bal95} with a 40\% covering factor, five-times solar metallicity, and \unit{250}{\kilo\meter~\reciprocal\second} turbulence. While these models reproduce the emission lines at slightly longer wavelengths, their behavior has not been explored in detail at the shorter wavelengths discussed here. The models likely suffer from inaccuracies due to high densities, ionizing fluxes, and temperatures in the emitting gas. These line identifications thus serve only as a guide until more extensive photoionization modeling is undertaken. In particular, some of the features correspond to lower-abundance elements such as argon, calcium, and sulfur which are unlikely to be strong.

Because all of the emission lines are highly blended, there is no straightforward way to isolate them and measure their strengths. Instead, Table~\ref{tab:ews} lists the emission EWs determined from the spline-fit Monte Carlo realizations for different regions of the composite spectra chosen to roughly separate the most prominent features. These divisions are plotted in Figures~\ref{fig:comp} and \ref{fig:nolowcomp} as vertical dot-dashed lines.  The reliability of these measurements is limited by the  power-law continuum fit against which they are referenced. Therefore most of the EWs likely underestimate the true emission flux if the power law is artificially high due to a pseudo-continuum of emission lines. Future studies may be able to more robustly separate and measure these emission lines by targeting AGNs with narrow emission lines (e.g., narrow-line Seyfert 1 galaxies), but all targets in our sample exhibit broad lines. The emission line strengths of the targets that contribute to the composites vary widely, as can be seen in the G140L spectra plotted in Figures~\ref{fig:specs1} and \ref{fig:specs2}. Some targets have nearly blazar-like continua with extremely weak emission features, while others have strong, prominent emission lines. This variation is illustrated in Figure~\ref{fig:ewplot}, which divides the sample of EWs for the contributing targets to each composite wavelength into quartiles.

\subsection{The \HeI\ 504~\AA\ Absorption Edge}

\begin{figure}
%final figure has been inserted%
\epsscale{1.15}
\plotone{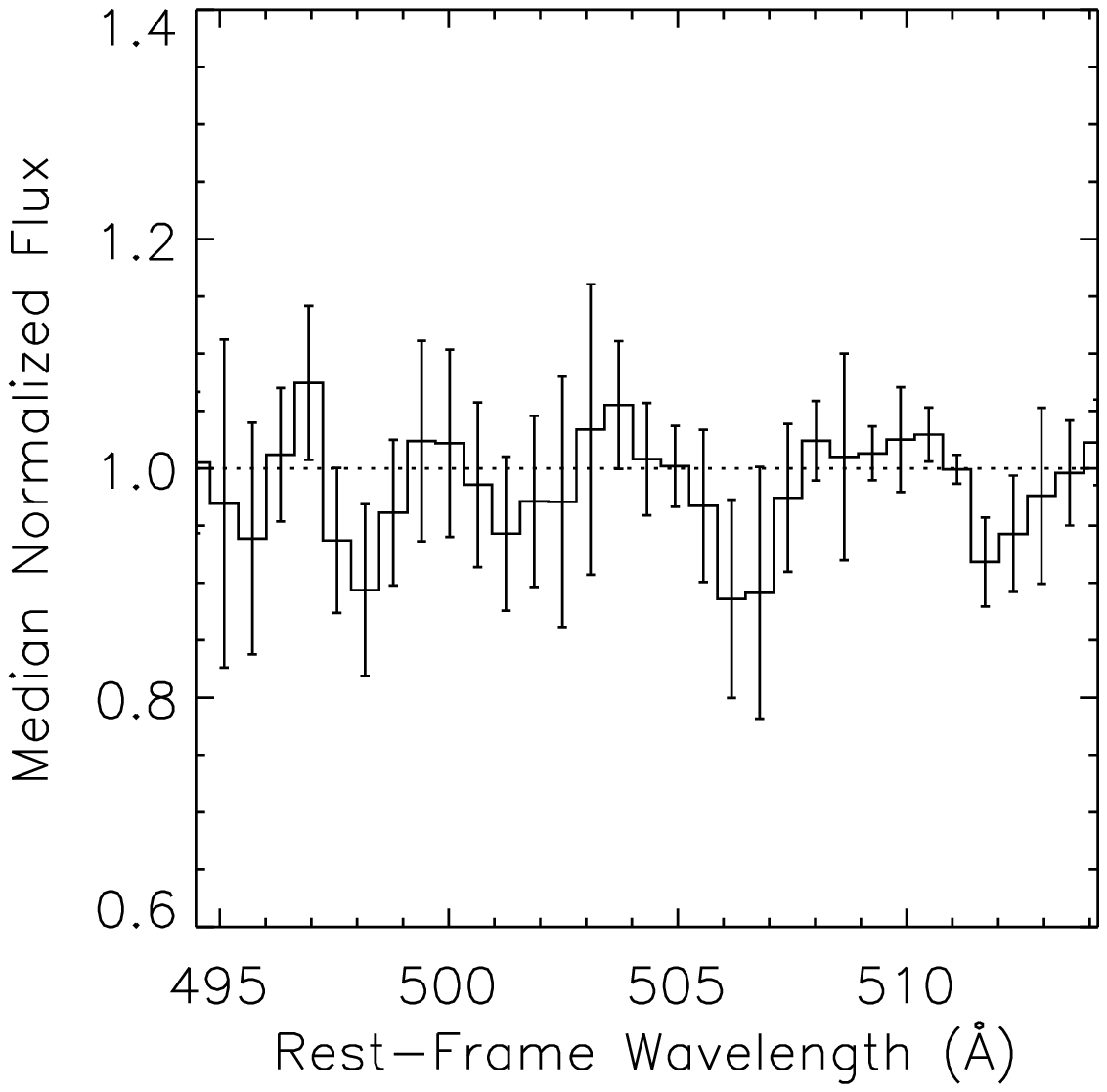}
\caption{A subsample composite of 5 AGN illustrating the non-detection ($\tau_{\rm HeI}<0.047$) of any edge-like spectral breaks due to \HeI\ absorption in the region around rest-frame 504~{\AA}.\label{fig:504edge}}
\end{figure}

%kolykhalov and sunyaev 1984, koratkar and blaes 1999, sincell and krolik 1997,  done and davis 2008, davis and laor 2011
%helium not there: done davis et al 2012, laor and davis 2014
A number of accretion disk atmosphere models have predicted the presence of absorption edges due to hydrogen, helium, and/or metal species in the spectra of AGN \citep[e.g.,][]{kol84,sin97,kor99,don08,dav11,don12,lao14}. Often, these models use atmospheres and winds analogous to those of hot stars,  applying stellar atmospheres codes such as \textsc{Tlusty} \citep{hub95,hub11} to generate spectra. The relative strengths and shapes of these edges vary greatly, ranging from broad spectral depressions to more distinctly edge-like features, depending on the details of the accretion disk models, such as the temperature profile of the disk atmosphere and the viewing angle of the observer. \citetalias{shu12a} and \citetalias{ste14} searched for and set upper limits on the continuum optical depth beyond the \HI\ 912~\AA\ edge ($\tau_{\rm HI}<0.01$) and the \HeI\ 504~\AA\ edge ($\tau_{\rm HeI}<0.1$). Our EUV sample does not probe any significant \HI\ or metal absorption edges likely to be present in an AGN spectrum, but the \HeI\ 504~\AA\ edge lies near the center of our wavelength coverage. 

To test for absorption edges in the region of 504~\AA, we constructed a composite spectrum from a subsample of five targets that have completely continuous, useable coverage with no masking in the $\sim 20~{\rm \AA}$ region surrounding the edge. We normalized this composite by the best-fit power law to remove the global slope. The flux in this spectra region is plotted in Figure~\ref{fig:504edge}. We set an optical depth limit by fitting the stacked flux to a model of pure \HeI\ continuum opacity using empirical photoionization cross-sections \citep{mar76}. This procedure yields a $2\sigma$ upper limit to the optical depth just beyond the edge, $\tau_{\rm HeI}<0.047$, corresponding to a column density of $\log N_{\rm HeI}<{15.8}$.

Despite the non-detection of a \HeI\ edge, there is weak evidence for \HeI\ emission longward of the edge, as shown by the measurements discussed in Section~\ref{sec:emission}. This detection is  at less than the $2\sigma$ level, however, and may be contaminated by other emission lines or by improper identification of the continuum flux with the power-law fit. If the \HeI\ edge is substantially smeared out or broadened, whether by inclination, disk atmospheric effects, or uncertainties in the redshifts used in our composite, the absorption would be difficult to identify amidst the numerous emission lines in this region. Nonetheless, there does not appear to be any region in which the \HeI\ optical depth could plausibly exceed $\tau_{\rm HeI}\approx 0.08$, corresponding to $ N_{\rm HeI}<10^{16}~{\rm cm^{-2}}$. It seems unlikely that \HeI\ opacity plays a significant role in the structure of AGN spectra.

\section{Summary and Conclusions}
\label{sec:conc}

In this paper, we have continued and updated our investigation of the mean AGN spectrum that we began in \citetalias{shu12a} and \citetalias{ste14}. We obtained 11 new COS/G140L spectra of AGNs at redshifts $1.45\leq z_{\rm AGN}\leq 2.14$ and combined their UV spectra with 9 existing medium-resolution COS spectra of AGNs to better characterize the typical spectral properties of AGNs in the rest-frame EUV. Using these data, we constructed composite spectra to aid in the determination of the typical spectral slope, the identification of the major emission lines, and the search for absorption edges predicted by some accretion disk models.

\noindent The major findings of this study are:
\begin{itemize}
\item Compared to  $700~{\rm \AA} <\lambda<900~{\rm \AA}$, the $450~{\rm \AA}<\lambda<770~{\rm \AA}$ region of AGN spectra exhibits a hardening spectral slope, $F_\nu\propto \nu^{-0.72\pm 0.26}$. Similar hardening  can be seen in the composites from \citet{tel02} and \citetalias{ste14}. Though this hardening may represent curvature of the continuum slope, it could represent a failure of the power-law model to separate continuum from a pseudo-continuum of emission-line flux.
\item The EUV spectral band contains a variety of prominent broad emission lines from ions of O, Ne, Mg, and other elements that are difficult to disentangle and identify. Possible identifications based upon the \textsc{Cloudy} models of \citet{mol14} were presented in Section~\ref{sec:emission} along with measurements of the strength of emission in different spectral regions. More detailed photoionization modeling and observations targeting AGNs with narrower emission lines are required to fully understand these features.
\item The \HeI\ 504~\AA\ absorption edge predicted by some accretion disk models is not detected. Our composite spectra limit the typical \HeI\ optical depth just below the edge to $\tau_{\rm HeI}<0.047$.

\end{itemize}

The sample of targets observed at high-S/N with EUV coverage below 500~\AA\ in the rest frame remains extremely small and is biased toward the most luminous sources. Nevertheless, COS offers the opportunity to expand this sample with AGNs at $z_{\rm AGN}\simeq 1.9-2.4$. It would also be useful to obtain NUV spectra necessary to characterize the intervening \HI\ absorption that introduces systematic error to measurements of AGN spectral features. This resource should be exploited while it is available, and it should be combined with photoionization modeling to understand the numerous emission features that complicate the interpretation of these spectra. Such studies are necessary steps toward understanding the processes that govern AGN as well as their effects on the IGM through their contributions to the UVB.

\acknowledgments

The authors thank Joshua Moloney for helpful discussions and assistance with emission line identification. Support for program number HST-GO-13302.01 was provided by NASA through a grant from the Space Telescope Science Institute. EMT acknowledges support from NASA Earth and Space Science Fellowship grant NNX14-AO18H.

{\it Facilities:} \facility{HST (COS, STIS)}.

\bibliographystyle{apj}
\bibliography{apj-jour,agnpaperbib}

\end{document}